\documentclass[10pt, conference]{IEEEtran}
\IEEEoverridecommandlockouts

\usepackage{cite}
\usepackage{amsmath,amssymb,amsfonts}
\usepackage{algorithm}
\usepackage{algpseudocode}
\usepackage{graphicx}   
\usepackage[T1]{fontenc}
\usepackage{subfigure}
\usepackage{adjustbox}
\usepackage{textcomp}
\usepackage{xcolor}
\usepackage{verbatim}
\usepackage{amsthm}
\usepackage[colorlinks=true, linkcolor=black, citecolor=blue, urlcolor=blue]{hyperref}

\def\BibTeX{{\rm B\kern-.05em{\sc i\kern-.025em b}\kern-.08em
    T\kern-.1667em\lower.7ex\hbox{E}\kern-.125emX}}

\begin{document}
\makeatletter
\newcommand*{\rom}[1]{\expandafter\@slowromancap\romannumeral #1@}
\makeatother

\makeatletter
\setkeys{Gin}{width=\ifdim\Gin@nat@width>\columnwidth
  \linewidth
\else
  \Gin@nat@width
\fi}
\makeatother

\title{Energy Saving for Cell-Free Massive MIMO Networks: A Multi-Agent Deep Reinforcement Learning Approach} 
\author{\IEEEauthorblockN{
Qichen Wang\IEEEauthorrefmark{1}, Keyu Li\IEEEauthorrefmark{1}, Ozan Alp Topal\IEEEauthorrefmark{1}, \"Ozlem Tu\u{g}fe Demir\IEEEauthorrefmark{2}, Mustafa Ozger\IEEEauthorrefmark{3}\IEEEauthorrefmark{1}, and Cicek Cavdar\IEEEauthorrefmark{1}
} 
\IEEEauthorblockA{\IEEEauthorrefmark{1}Department of Communication Systems, KTH Royal Institute of Technology, Sweden (\{qichenw, keyul, oatopal, cavdar\}@kth.se)}
\IEEEauthorblockA{\IEEEauthorrefmark{2}Department of Electrical and Electronics Engineering, Bilkent University, Turkiye  (ozlemtugfedemir@bilkent.edu.tr) }
\IEEEauthorblockA{\IEEEauthorrefmark{3}Department of Electronic Systems, Aalborg University, Denmark  (mozger@es.aau.dk) }
}


\maketitle
\begin{abstract}
This paper focuses on energy savings in downlink operation of cell-free massive MIMO (CF mMIMO) networks under dynamic traffic conditions. We propose a multi-agent deep reinforcement learning (MADRL) algorithm that enables each access point (AP) to autonomously control antenna re-configuration and advanced sleep mode (ASM) selection. After the training process, the proposed framework operates in a fully distributed manner, eliminating the need for centralized control and allowing each AP to dynamically adjust to real-time traffic fluctuations. Simulation results show that the proposed algorithm reduces power consumption (PC) by 56.23\% compared to systems without any energy-saving scheme and by 30.12\% relative to a non-learning mechanism that only utilizes the lightest sleep mode, with only a slight increase in drop ratio. Moreover, compared to the widely used deep Q-network (DQN) algorithm, it achieves a similar PC level but with a significantly lower drop ratio. 
\end{abstract}


\section{Introduction}

Cell-free massive MIMO (CF mMIMO) has emerged as a promising architecture for future mobile communication systems, primarily due to its ability to provide almost uniformly high quality of service (QoS) to user equipments (UEs) across the network \cite{interdonato2019ubiquitous}. In a typical CF mMIMO setup, a large number of distributed access points (APs) jointly serve a smaller number of UEs through coherent transmission. To accommodate the increasing demand and growing number of UEs, this architecture relies on an ultra-dense deployment of APs. However, such deployment inevitably incurs substantial energy consumption \cite{van2020optimal}, primarily due to the increased number of APs and the overhead of synchronization and control signaling across these densely deployed APs. 

Several studies have been conducted to address the energy efficiency challenges in CF mMIMO systems.
The works in \cite{mohammed2024cell, shi2024joint} investigate AP selection as a means to improve energy efficiency, where UEs are served by subsets of APs, chosen based on channel conditions or power thresholds, rather than by all available APs. Similarly, \cite{demir2024energy} focuses on AP clustering and simple AP activation strategies to enhance energy efficiency.
In addition, AP sleep mode has been widely studied as another line of research. The studies in \cite{9149862,10380322} propose adaptive deactivation of underutilized APs, with \cite{10380322} further enhancing the approach by incorporating transmit power optimization. Meanwhile, \cite{riera2023energy} introduces a multi-level sleep mode framework and demonstrates the effectiveness of coordinated sleep scheduling. As a complementary technique, antenna-level optimization has also been considered. \cite{10942847} introduces a joint scheme of power allocation and antenna deactivation for energy-efficient CF mMIMO under wireless fronthaul.

The current research on energy-efficient CF mMIMO systems largely relies on snapshot-based settings, where all UE arrivals are assumed to be known at a single time instant. These static approaches overlook temporal dynamics in decision-making, and thus cannot effectively model realistic traffic variations. In our previous work \cite{10624787}, dynamic traffic arrivals were introduced but only in a cellular network context. Regarding CF mMIMO, determining antenna element configurations and managing AP sleep modes under dynamic traffic conditions remains an open challenge. In this paper, we propose a multi-agent deep reinforcement learning (MADRL) algorithm that enables APs to jointly optimize antenna re-configuration and advanced sleep mode (ASM) selection under dynamic traffic.
The main contributions of this paper are outlined as follows:
\begin{itemize}
\item We develop a comprehensive CF mMIMO simulation framework that integrates empirically derived mobile traffic patterns from deep packet inspection (DPI) data of a Swedish operator, thereby enabling realistic and dynamic evaluation of system performance under practical traffic conditions.
\item We propose a multi-agent proximal policy optimization (MAPPO)-based algorithm that jointly manages antenna re-configuration and ASMs selection of APs, with the objective of minimizing power consumption (PC) while maintaining the data rate satisfaction. 
\item Simulation results show that the proposed MAPPO-based algorithm achieves substantial energy savings compared to non-learning baselines with simple energy-saving mechanisms. Moreover, it outperforms the deep Q-learning network (DQN) algorithm by learning better traffic-sensitive policies that achieve a lower drop ratio under the same PC. 
\end{itemize}


\section{System Model}
\begin{figure}[h]
    \centering
    \includegraphics[width=0.9\linewidth]{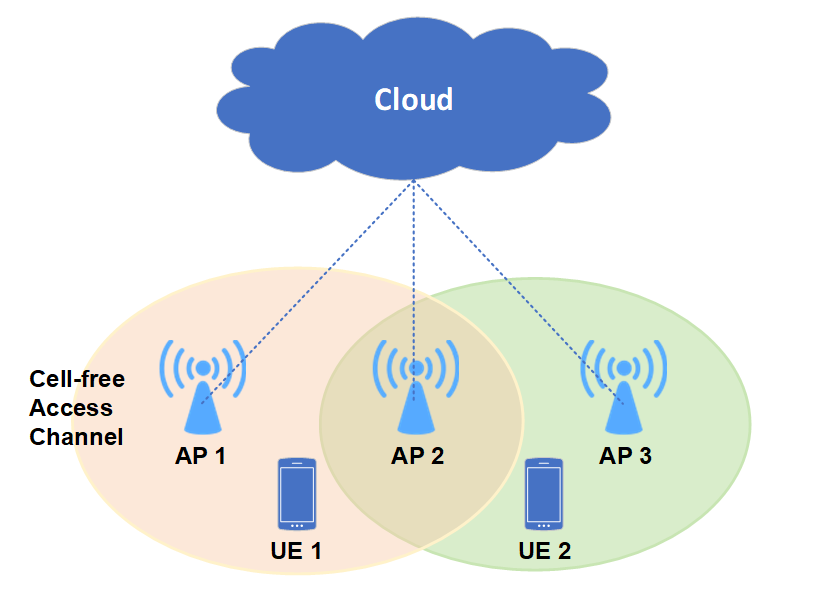}
    \caption{CF MIMO system model.}
    \label{fig:system model}
\end{figure}

We focus on the downlink of a CF mMIMO system, as illustrated in Fig. \ref{fig:system model}. In our system model, $L$ APs, each equipped with $M_{\max}$ antennas, are deployed at fixed locations and connected to a centralized cloud via fronthaul links. Within the coverage area, UEs with single-antenna arrive randomly at various locations according to an arrival rate $\lambda$. When a UE arrives, it selects a subset of APs with the strongest channel gains to which it sends its service request. In response to dynamic traffic conditions, APs can implement adaptive energy-saving strategies, including turning antennas on and off, and transitions among multiple sleep modes.
\vspace{-4mm}
\vspace{-4mm}
\subsection{Channel Model}
The channel between AP $l$ and UE $k$ is characterized by a large-scale fading coefficient $\beta_{l,k}$, capturing path loss and shadowing. The small-scale fading is assumed to follow independent and identically distributed Rayleigh fading.

The coherence block has $\tau_c$ symbols, of which $\tau_p$ are used for uplink channel estimation and  $\tau_c - \tau_p$ are used for downlink data. During the uplink training phase, each UE transmits a pilot of length $\tau_p$. In a large network with $K$ UEs (where $K > \tau_p$), it is not possible to assign orthogonal pilots to all the UEs. Instead, some UEs can share the same pilot sequence. The set of UE indices that share the same pilot as UE $k$ is denoted by $\mathcal{P}_k$. Each AP obtains the channel estimate of UEs, where the average gain of the channel estimate of UE $k$ at AP $l$ is denoted by $\chi_{l,k}$. 

To keep the architecture scalable, we follow a user-centric AP-UE association: each UE~$k$ is jointly served only by a subset of APs, $\mathcal{M}_k \subseteq \{1,\dots,L\}$, instead of by the whole network, $L$ denoting the total number of APs. We rank the large-scale fading coefficients in descending order and construct the cluster as
\begin{equation}
\sum_{i=1}^{|\mathcal{M}_k|} \beta_{\ell_i,k} \ge 0.9 \sum_{l=1}^L \beta_{l,k},
\label{eq:adaptive-cluster}
\end{equation}
where $|\mathcal{M}_k|$ is the adaptive cluster size and $l_i$'s are the indices of the APs serving UE $k$. This $90\%$-gain rule~\cite{demir2021foundations} guarantees that at least $90\%$ of the available large-scale energy reaches UE~$k$ while dramatically shrinking the signaling footprint and precoding dimension.

Each AP precodes and transmits data streams intended for the UEs in its serving set. Let $\mathcal{K}_l \subseteq \{1,\ldots,K\}$ denote the set of UEs served by AP $l$. The transmitted signal from AP $l$ is given by
\begin{equation}
   \mathbf{x}_l = \sum_{k \in \mathcal{K}_l} \sqrt{p_{l,k}}\mathbf{w}_{l,k}\varsigma_k,
\end{equation}
where $\varsigma_k$ is the data symbol intended for UE $k$ with $\mathbb{E}[|\varsigma_k|^2] = 1$, $\mathbf{w}_{l,k} \in \mathbb{C}^{m_l}$ is the precoding vector, where $m_l$ denotes the number of activated antennas at AP $l$, and $p_{l,k}$ is the transmit power allocation coefficient, determined by our power allocation policy (see Section~\ref{sec:power-allocation}).

To mitigate downlink interference in a distributed manner, we adopt the local protective partial zero-forcing (PPZF) precoding scheme~\cite{9069486} due to its balanced complexity-performance trade-off. 

The received signal at UE $k$ consists of the desired component, coherent interference due to pilot contamination, and non-coherent interference plus noise. Thus, the effective signal-to-interference-plus-noise ratio (SINR) with PPZF precoding is given by \eqref{eq:SINR_PPZF}, where $\tau^{\text{str}}_l\leq\tau_p$ represents the number of pilot signals for the strong-channel UEs at AP $l$. The indicator $\delta_{l,k}$ specifies whether UE $k$ belongs to the interference-suppression set of AP $l$, while the set $\mathcal{P}_k$ comprises the UEs sharing the same pilot sequence with UE $k$, thereby capturing the effect of pilot contamination. The $\sigma^2$ is the noise variance.

\begin{figure*}[t]
  \begin{equation}
\text{SINR}_k^{\text{PPZF}} =
 \frac{
\left( \sum_{l=1}^L \sqrt{(m_l - \tau^{\text{str}}_l) p_{l,k} \chi_{l,k}} \right)^2
}{
\sum_{t \in \mathcal{P}_k \setminus \{k\}} \left( \sum_{l=1}^L \sqrt{(m_l - \tau^{\text{str}}_l) p_{l,t} \chi_{l,k}} \right)^2 
 \quad + \sum_{t=1}^{K} \sum_{l=1}^{L} p_{l,t} \left( \beta_{l,k} - \delta_{l,k} \chi_{l,k} \right) + \sigma^2
}.
\label{eq:SINR_PPZF}
\end{equation}
\hrulefill
\end{figure*}

Finally, the achievable downlink data rate for UE~$k$ is computed as:
\begin{equation}
r_k = \left(\frac{\tau_c - \tau_p}{\tau_c} \right) B \log_2\left(1 + \text{SINR}_k^{\text{PPZF}}\right),
\end{equation}
where $B$ is the system bandwidth.


\subsection{Power Allocation}
\label{sec:power-allocation}
For AP $l$, the total transmit power is  defined as $p_l = m_l p_a$, where $p_a$ is the average transmit power per antenna. Equivalently, it can be expressed as $p_l = \sum_{k \in \mathcal{K}_l} p_{l,k}$. The power allocated to UE~$k \in \mathcal{K}_l$ according to \cite{demir2021foundations} is given by:
\begin{equation}
p_{l,k} = p_l \cdot \frac{\chi_{l,k}}{\sum_{j \in \mathcal{K}_l} \chi_{l,j}}
\end{equation}
where $\chi_{l,k}$ reflects the quality of the channel estimate between AP $l$ and UE $k$, thereby favoring UEs with stronger channels.

\subsection{Advanced Sleep Modes}
ASMs enable the AP to gradually enter deeper sleep modes by systematically deactivating hardware components such as the radio frequency (RF) module and the power amplifier (PA). Although higher sleep modes yield significant energy savings, they involve a trade-off in the form of increased wake-up latency. According to  \cite{10437599,lozano2025kairos}, four sleep modes (SM 0–3) are defined based on the associated wake-up latency, each with a corresponding PC discount factor, as summarized in Table \ref{tab:sms}. Among them, SM 0 denotes the active state.
\begin{table}[H]
\caption{Advanced sleep modes \cite{lozano2025kairos}.} \label{tab:sms}
\begin{centering}
\begin{tabular}{|l|c|c|c|c|}
\hline 
Sleep mode $s$ & 0 & 1 & 2 & 3\tabularnewline
\hline 
Wake-up latency $\Delta_s$ & $0$\,$\mu$s & $37$\,$\mu$s & $500$\,$\mu$s & $5000$\,$\mu$s\tabularnewline
\hline 
PC discount factor $\eta_s$ & $1$ & $0.675$ & $0.55$ & $0.23$\tabularnewline
\hline 
\end{tabular}
\par\end{centering}
\end{table}

\subsection{Power Consumption Model}\label{sec:pwr_model}
We adopt the functional split~7.2 as in open radio access network (O-RAN) architecture. 
All RF, filtering, fast Fourier transform (FFT) / inverse FFT (iFFT), and precoding operations are executed at the AP side, while modulation, coding, and higher-layer processing are carried out in a pool of general-purpose processors (GPPs)
in the cloud. Consequently, the network PC naturally splits into an AP part and a cloud part:

\begin{align}
P_{\text{net}}
      &=\underbrace{\sum_{l=1}^{L} P_l}_{\text{AP-side}}
        +\;\underbrace{P_{\text{cloud}}}_{\text{cloud-side}} .
\end{align}

\subsubsection{AP–side power}
For AP~$l$ ($l=1,\dots,L$), we follow the generic model in
\cite{10942847}.  The total instantaneous power is
\begin{equation}\label{eq:AP_total}
P_l =
m_lP_{\mathrm{st}}+
\Delta_{\mathrm{tr}}\!\sum_{k\in \mathcal{K}_l} p_{l,k}+
\left(P^{\mathrm{proc}}_{0}+
                 \Delta^{\mathrm{proc}}_{\text{AP}}\,
            \frac{C_{\text{AP},l}}{C^{\text{max}}_{\text{AP}}}\right),
\end{equation}
where $P_{\mathrm{st}}$ is the hardware-dependent static PC and the slope $\Delta_{\mathrm{tr}}$ models the load-dependent transmit power. The two terms in parentheses account for the processing power, where $P^{\mathrm{proc}}_{0}$ is the idle processing power per AP, and $\Delta^{\mathrm{proc}}_{\text{AP}}$ is the slope of the load-dependent processing PC for AP $l$. Here, $C_{\text{AP},l}$ denotes the processing utilization in giga-operations per second (GOPS) for AP $l$, and $C_{\text{AP}}^{\text{max}}$ represents the maximum processing capacity.
When the AP enters sleep mode
$s_l\in\{1,2,3\}$, its transmit power is set to zero, while the remaining terms are scaled down by empirical factors $\eta_{s_l}$
as proposed in~\cite{peesapati2021analytical}. This leads to the expression for the PC in sleep mode: $P^{\text{sleep}}_l=\eta_{s_l}P_l$.

\subsubsection{Cloud–side power}
The cloud PC can be given as
\begin{equation}\label{eq:P_cloud}
P_{\text{cloud}} =
P^{\text{fixed}}
+\frac{1}{\sigma_{\text{cool}}}\!
 \Bigl(P^{\text{comp}}_0
      +\Delta^{\text{proc}}_{\text{GPP}}\,
       \frac{C_{\text{GPP}}}{C^{\text{max}}_{\text{GPP}}}\Bigr),
\end{equation}
where $P^{\text{fixed}}$ denotes the load-independent fixed PC. The two terms in parentheses correspond to the processing power, adjusted by the cooling
efficiency $\sigma_{\text{cool}}$. Here, $P^{\text{comp}}_0$ is the idle processing power, and the remaining parameters are defined analogously to those on the AP side.

The detailed definitions of $C_{\text{AP},l}$ and $C_{\text{GPP}}$ can be found in \cite{10942847}.

\subsection{Traffic Model}
\label{sec:traffic}
We process DPI data from a mobile operator to build a realistic traffic model, 
with detailed processing methodology described in \cite{tianzhang2023mobile}. Traffic flows are categorized into three service classes $z \in \{\text{delay-stringent},~ \text{delay-sensitive},~ \text{delay-tolerant}\}$, with 3GPP packet delay budgets of 50\,ms, 100\,ms, and 150\,ms, respectively.

For each service class $z$, traffic flows are aggregated into 20-minute intervals, 
averaged over a week, and expressed as the temporal-spatial traffic density 
$\kappa_{z,t}\,[\text{Mbit}\cdot \text{s}^{-1}\cdot \text{km}^{-2}]$ for timestep $t$. This yields a time-varying traffic density profile, which is mapped to a dynamic arrival rate for UEs. Assuming that each UE initiates a single traffic flow of identical size $x^{\text{max}}$ (Mb), the arrivals are modeled as a space-homogeneous Poisson process with time-dependent mean:
\begin{equation}
\lambda_{z,t} = \frac{\kappa_{z,t} A \Delta t}{x^{\text{max}}},
\label{eq:arrival-rate}
\end{equation}
where $A$ denotes the area size (km$^2$), $\Delta t$ the simulation step duration, and $x^{\text{max}}$ the demand size (Mb). Hence, UE arrivals follow a non-stationary Poisson process, directly reflecting the temporal variations of the traffic load.

Each UE is assigned a delay budget $D^{\text{max}}_k$ according to its service class. Its demand decreases with the achieved data rate, while its delay decreases over time. A UE departs from the system when either its demand is fully served or its delay budget expires. At departure, the remaining demand and delay are denoted $x_{k}^{\text{rem}}$ and $D_{k}^{\text{rem}}$, respectively. We define the average required rate as $r^{\text{req}}_{k}=\frac{x^{\text{max}}}{D^{\text{max}}_k}$ and the average achieved rate as $\displaystyle  r^{\text{ach}}_{k}=\frac{x^{\text{max}}-x_{k}^{\text{rem}}}{D^{\text{max}}_k-D_{k}^{\text{rem}}}$.  Their ratio is given by $\displaystyle \rho_k=\frac{r^{\text{ach}}_{k}}{r^{\text{req}}_{k}}$. When the achieved rate is below the required rate ($r^{\text{ach}}_k<r^{\text{req}}_k$), the drop ratio can be expressed as $\displaystyle  1-\rho_k=\frac{x_{k}^{\text{rem}}}{x^{\text{max}}}$. 

\subsection{Problem Formulation}
Our objective is to minimize the network PC while guaranteeing a low average drop ratio. The problem can be formulated as:
\begin{align}
\min_{s_{l},\,m_{l},\,\forall l} &\quad P_{\text{net}} \label{Problem}\\
\text{s.t.}
&\quad \frac{1}{K}\sum^K_{k=1}\frac{x_{k}^{\text{rem}}}{x^{\text{max}}}\leq\delta_{\text{drop}}, &\forall k,\tag{\ref{Problem}{a}} \label{Problema}\\
&\quad m_l \in \left\{0, 1, \dots, M_{\text{max}} \right\}, &\forall l,\tag{\ref{Problem}{b}} \label{Problemb}\\
&\quad \tau^{\text{str}}_l\in\{0,1,\dots,m_l-1\}, &\forall l,\tag{\ref{Problem}{c}}\label{Problemc}\\
&\quad s_l\in\{0,1,2,3\}, &\forall l.\tag{\ref{Problem}{d}}\label{Problemd}
\end{align}
Constraint (\ref{Problema}) enforces that the average drop ratio remains below the threshold value $\delta_{\text{drop}}$. Constraint (\ref{Problemb}) ensures that the number of active antennas at each AP is an integer variable that does not exceed the equipped antennas. Constraint (\ref{Problemc}) guarantees a non-zero effective SINR for PPZF precoding as shown in \eqref{eq:SINR_PPZF}. Constraint (\ref{Problemd}) regulates the available sleep modes.

\section{Multi-Agent PPO-Based Algorithm}
In this section, we propose a MAPPO-based resource allocation algorithm to jointly optimize antenna re-configuration and ASMs selection. Each AP is modeled as an agent with its own actor network, while a centralized critic deployed in the cloud leverages global information to guide training. The framework follows the centralized training and decentralized execution (CTDE) paradigm: actor and critic networks are jointly trained in the cloud, and only the actor networks are deployed at the APs for fully decentralized decision-making during execution.

\subsection{Markov Decision Process Model}
We formulate the multi-agent resource allocation problem as a Markov decision process (MDP), represented by the tuple ⟨$\mathcal{S}, \mathcal{A}, P, R, \gamma$⟩:
\begin{itemize}
    \item \textbf{State space $\mathcal{S}$:} At timestep $t$, each agent observes partial local information $o_t$, including its own configuration (e.g., PC, number of activated antennas, sleep mode) and aggregate UE statistics (e.g., total demand and achieved rate), along with neighboring AP states. Relying on such partial and varying observations makes independent critics suffer from non-stationarity, as the environment dynamics change with concurrently learning agents. To address this, we adopt a centralized critic with access to the global state $s_t$ (e.g., total network PC, UE arrival rate, average drop ratio, and delay ratio), which provides a more stationary learning signal and a consistent value estimation that stabilizes training.
    \item \textbf{Action space $\mathcal{A}$:} The action space is defined as $\mathcal{A} = \mathcal{A}_m \times \mathcal{A}_s$, where $\mathcal{A}_m = \{-1,0,+1\}$ denotes antenna re-configuration actions (activating or deactivating one antenna at a time), and $\mathcal{A}_s = \{0,1,2,3\}$ represents available sleep mode choices.
    \item \textbf{Reward function $R$:} For UE $k$, we introduce the rate satisfaction (RS) score function as:
\begin{align}
&~\xi_{k}=\begin{cases}
\rho_{k}-1, & \rho_{k}<1,\\
\phi\left(1-\frac{1}{\rho_{k}}\right), & \rho_{k}\ge 1.
\end{cases}
\end{align}

If $\rho_{k}<1$, the achieved rate is insufficient to meet the demand, and $\xi_{k}=\rho_{k}-1$ is negative, serving as a penalty term, and its magnitude directly equals the drop ratio. Conversely, when $\rho_{k}\ge 1$, the UE’s demand can be fully met. In this case, a non-negative reward is assigned according to $\xi_{k}=\phi\left(1-\frac{1}{\rho_{k}}\right)$, with $\frac{1}{\rho_{k}}$ quantifying the proportion of the experienced delay relative to the delay budget. The coefficient $\phi$ acts as an attenuation factor scaling the positive component of $\xi_k$. As shown in Fig.~\ref{fig:phi}, a smaller $\phi$ yields a slower growth of the positive term with $\rho$, thus keeping the optimization biased toward reducing data drops. 
\vspace{-2mm}
\begin{figure}[h]
    \centering
    \includegraphics[width=0.75\linewidth,trim=5 5 45 40,clip]{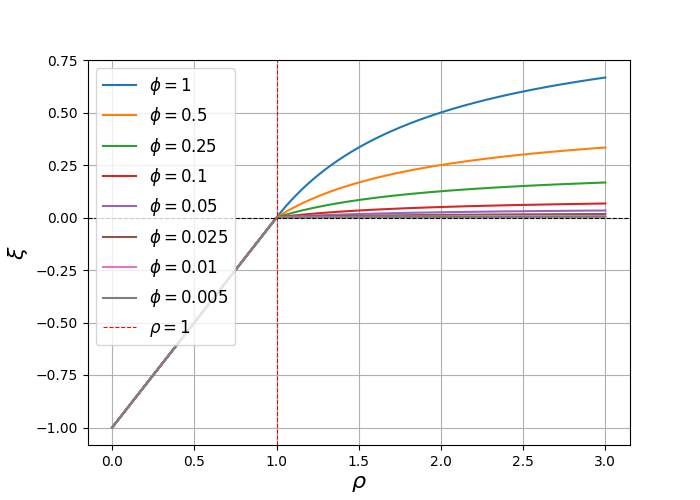}
    \vspace{-3mm}
    \caption{$\xi$ vs. $\rho$ for different $\phi$ values.}
    \label{fig:phi}
    \vspace{-3mm}
\end{figure}

Defining a global reward encourages agents to cooperate by optimizing a shared objective, thereby mitigating non-stationarity. This is conceptually similar to the role of a centralized critic, which leverages global information to provide consistent learning signals. Accordingly, we design the global reward at timestep $t$ as:
\begin{equation}
    R=w_{\text{qos}}\frac{1}{K}\sum_{k=1}^{K}\xi_{k}-w_{\text{pc}}P_{\text{net}},\label{eq:reward}
\end{equation}
where $w_{\text{rs}}$ and $w_{\text{pc}}$ denote the weights to balance RS score and PC. $K_t$ is the total number of UEs at timestep $t$.
\item \textbf{Transition probability $P$ and discount factor $\gamma$:} The environment dynamics, including traffic arrivals, UE associations, and channel variations, determine the transition from $s_t$ to $s_{t+1}$ under joint actions $\mathbf{a}_t$, while a discount factor $\gamma \in (0,1]$ balances immediate and future rewards.
\end{itemize}

\subsection{Learning Process}
At timestep $t$, the actor network with parameter $\theta$ samples an action $a_t$ from a probability distribution, $\pi_\theta\left(a_t\mid o_t\right)$, generated over its action space. The critic network with parameter $\varphi$ estimates the value of the current state $\hat{V}_\varphi(s_{t})$ and the value of the next state, $\hat{V}_\varphi(s_{t+1})$, resulting from selected action. Given the collected values, the temporal-difference (TD) error can be computed as:
\begin{equation}
\tilde{\delta}_t = R_t + \gamma\hat{V}_\varphi(s_{t+1}) - \hat{V}_\varphi(s_{t}).
\end{equation}
Then, the actor network with parameter $\theta$ is updated by maximizing the following objective function:
\begin{equation}
    \begin{split}
        L\left(\theta\right)=&\mathbb{E}\left[\min\left(r_t(\theta)\hat{A}_{t}, \right.\right.\left.\left.\text{clip}\left(r_t(\theta),1-\varepsilon,1+\varepsilon\right)\hat{A}_{t}\right)\right] \\&+ c_{e}H\left(\pi_{\theta}\mid o_{t}\right),\label{eq:policy-loss}
    \end{split}
\end{equation}
where $\hat{A}_{t}=\sum_{k=0}^{T-t}(\gamma\psi)^k\tilde{\delta}_{t+k}$ is the advantage function approximated by generalized advantage estimation (GAE), the parameter $\psi$ to balance bias and variance, $r_t(\theta)=\frac{\pi_\theta\left(a_t\mid o_t\right)}{\pi_{\theta_{\text{old}}}\left(a_{t}\mid o_t\right)}$ denotes the ratio of action selection probabilities under the current policy $\pi_\theta$ relative to the previous policy $\pi_{\theta_\text{old}}$, $c_eH\left(\pi_{\theta}\mid o_t\right)$ presents the entropy bonus. The clip function, by restricting the probability ratio between the new and old policies, is a key mechanism that enables PPO to maintain stable convergence: it prevents overly large updates, reduces training oscillations, and ensures smoother convergence. The advantage becomes particularly pronounced in multi-agent environments, where agents can easily affect one another and induce environmental instability. In such settings, the clip function suppresses overly aggressive updates, so that enhances cooperative stability among agents.

The critic network can be updated by minimizing the Huber loss function:
\begin{equation}
    L_{V}\left(\varphi\right)=\mathbb{E}\left[L_{\epsilon}^{\text{Hb}}\left(\tilde{\delta}_{t}\right)\right],\label{eq:value-loss}
\end{equation}
with
\begin{equation}
    L_\epsilon^\text{Hb}\left(e\right)=\begin{cases}
 		\frac{1}{2}e^2, & \left|e\right|\le\epsilon,\\
 		\epsilon\left(\left|e\right|-\frac{1}{2}\epsilon\right), & \left|e\right|>\epsilon,
 	\end{cases}
\end{equation}
where $\epsilon$ is the threshold parameter. The Huber loss function integrates the strengths of both mean squared error (MSE) and mean absolute error (MAE), offering a balance between sensitivity to outliers and stable gradient behavior.

\section{Numerical Results}
Our simulation area is a square region of size $1 \,\text{km} \times 1 \,\text{km}$, where $L = 25$ APs are deployed in a regular $5 \times 5$ grid. 
Each AP is equipped with $M_{\max} = 8$ antennas, where each antenna operates with an average transmit power of $p_a = 250$\,mW. The large-scale fading coefficient $\beta_{l,k}$  follows the 3GPP urban microcell (UMi) non-line-of-sight (NLOS) models \cite{3gpp38901}. The system operates over a bandwidth of $B = 20$\,MHz centered at carrier frequency $f_c = 5$\,GHz, with a pilot length of $\tau_p = 7$. The drop ratio threshold is set to $\delta_{\text{drop}}=0.1\%$. The UEs are generated uniformly within the area following the traffic model  in Section \ref{sec:traffic}, each with a demand size of $x^{\text{max}}=1.5$\,Mbits.

MAPPO agents are trained for 200 episodes, each simulating one week, with timestep $\Delta t=1$\,ms. Actions are selected every 20 timesteps ($20$\,ms), consistent with the default periodicity of synchronization signal block transmission~\cite{peesapati2021analytical}. The main hyperparameters are listed in Table \ref{tab:mappo-hyperparams}.
In the reward function, $w_{\text{rs}}=60$ (with attenuation $\phi=5\times10^{-3}$) and $w_{\text{pc}}=0.4$.
\begin{table}[h]
\centering
\caption{Hyperparameters used for MAPPO training.}
\label{tab:mappo-hyperparams}
\begin{tabular}{|l|c|}
\hline
\textbf{Parameter} & \textbf{Value} \\
\hline
Discount factor $\gamma$ & $0.99$ \\
Entropy coefficient $c_e$  & $0.01$ \\
Actor learning rate $\eta_\pi$  & $5 \times 10^{-4}$ \\
Critic learning rate $\eta_v$  & $5 \times 10^{-4}$ \\
PPO epochs  & $10$ \\
Mini-batches  & $32$ \\
Clip parameter $\varepsilon$ & $0.2$ \\
GAE parameter $\psi$ & $0.95$ \\
Huber loss parameter $\epsilon$ & $10$ \\
\hline
\end{tabular}
\end{table}

We compare the performance of our proposed MAPPO-based algorithm with two non-learning and one learning-based baselines. The first baseline, \emph{Always-on}, keeps all APs active and all antennas permanently turned on without any energy-saving mechanisms. 
The second, \emph{dynamic antenna configuration with SM1} (\emph{DAC-SM1}), switches an AP to SM1 when idle and reactivates it upon UE association, while antenna activation is stepwise adjusted using dual thresholds on the ratio of total achieved to demand rate of its connected UEs (deactivating one antenna if the ratio exceeds 55, activating one if it falls below 45).
As a representative learning-based benchmark, we also include DQN, a widely adopted DRL algorithm. 
\vspace{-4mm}
\begin{figure}[h]
    \centering
    \includegraphics[width=1.0\linewidth, trim=100 50 55 40,clip]{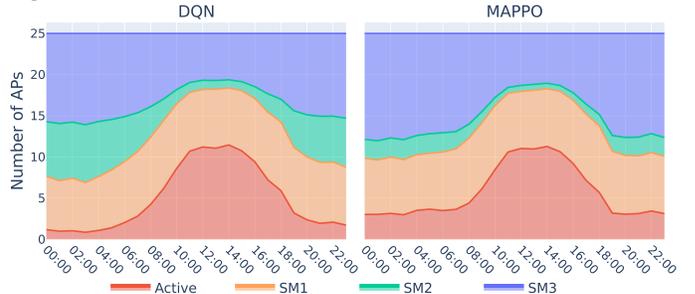}
    \vspace{-5mm}
    \caption{Time-varying number of APs in different sleep modes.}
    \label{fig:sm}
\end{figure}
\vspace{-2mm}
\begin{figure*}[t]
\begin{minipage}{0.31\textwidth}
    \includegraphics[width=\linewidth, trim=15 25 200 90, clip]{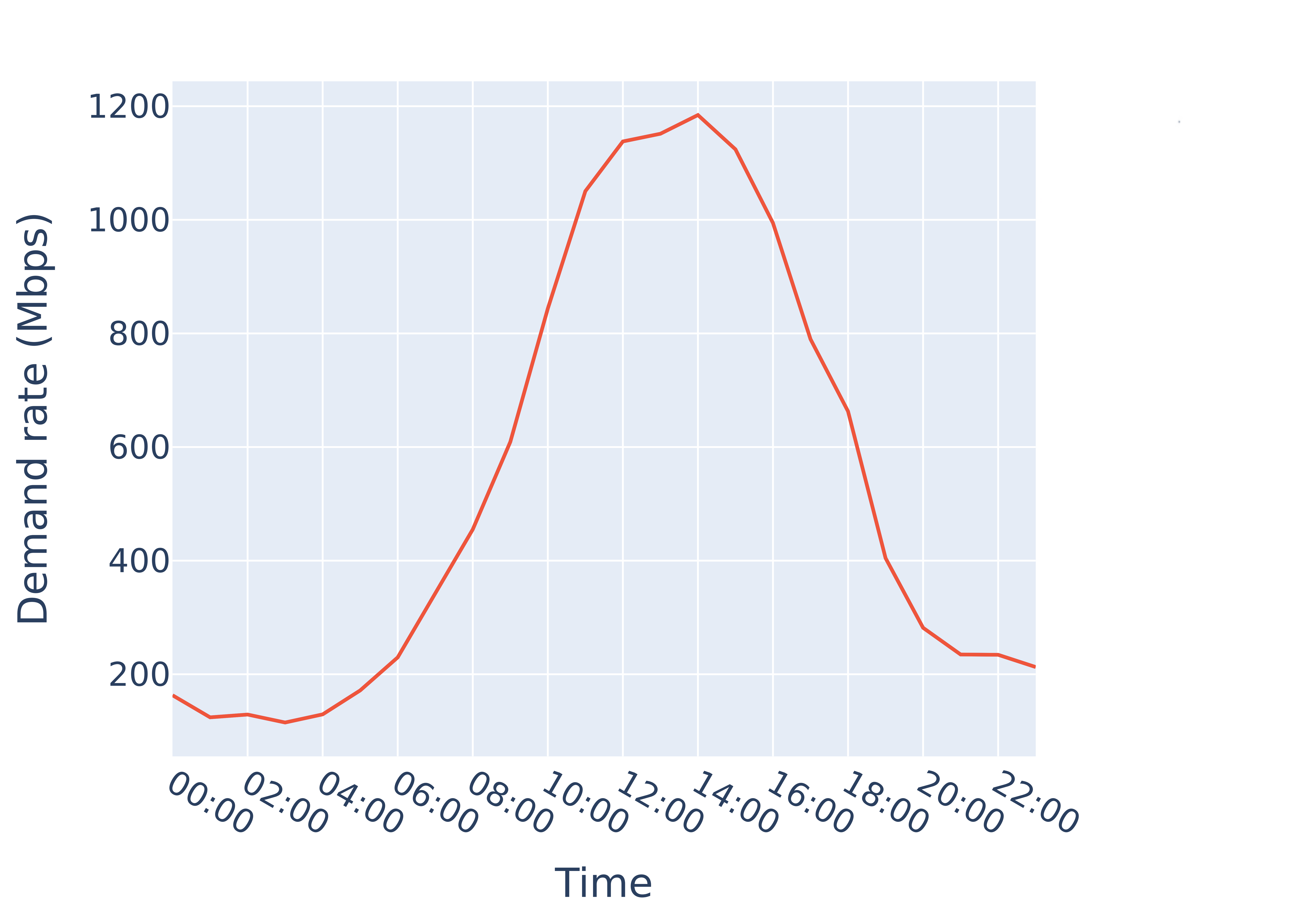}
    \caption{Total demand rate. \label{fig:traffic}}
\end{minipage}
\begin{minipage}{0.34\textwidth}
    \centering
    \includegraphics[width=\linewidth, trim=20 25 30 90, clip]{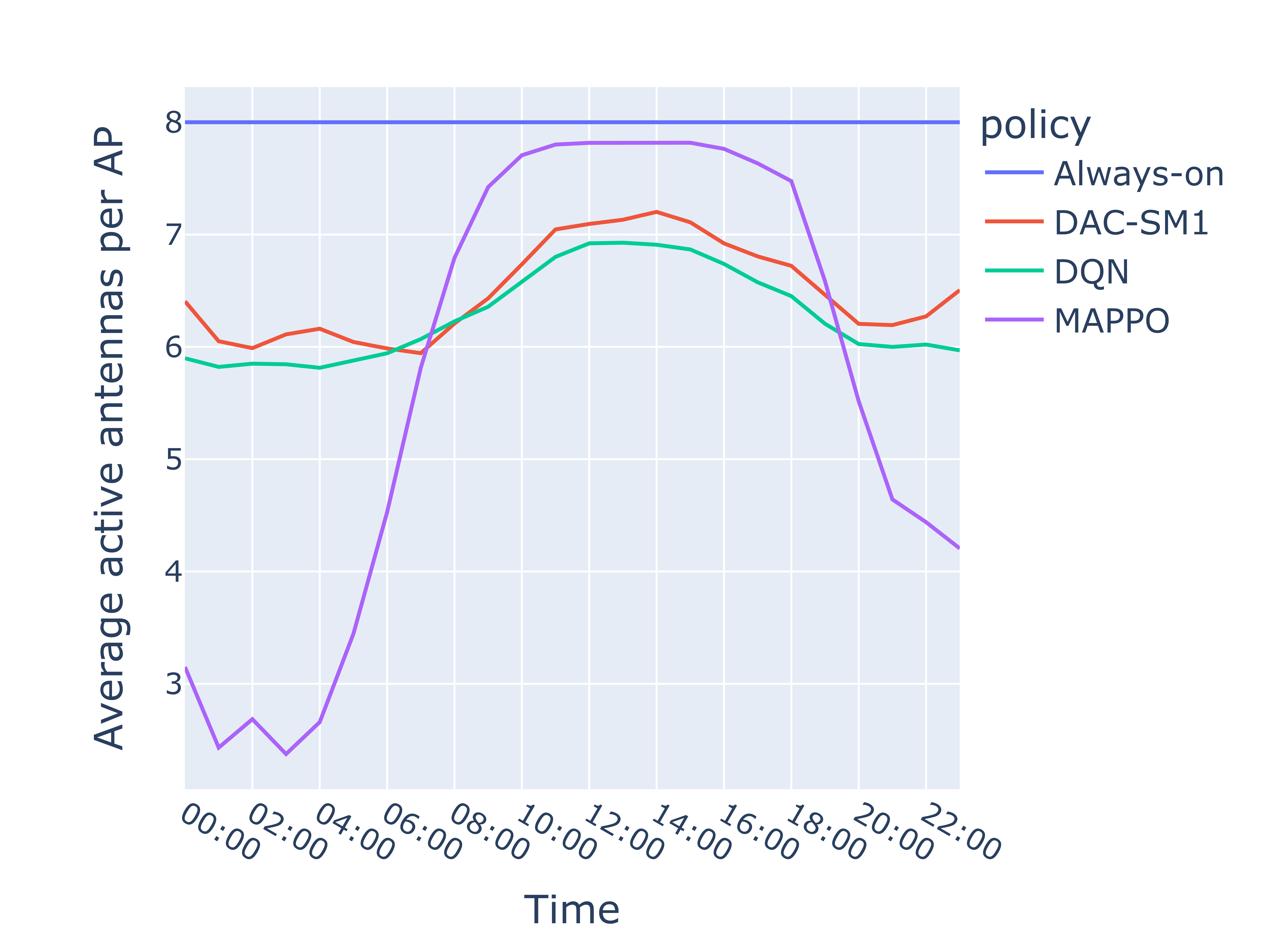}
    \caption{Average number of active antennas per AP. \label{fig:ant}}
\end{minipage}
\begin{minipage}{0.32\textwidth}
    \centering
    \includegraphics[width=\linewidth, trim=8 10 8 10, clip]{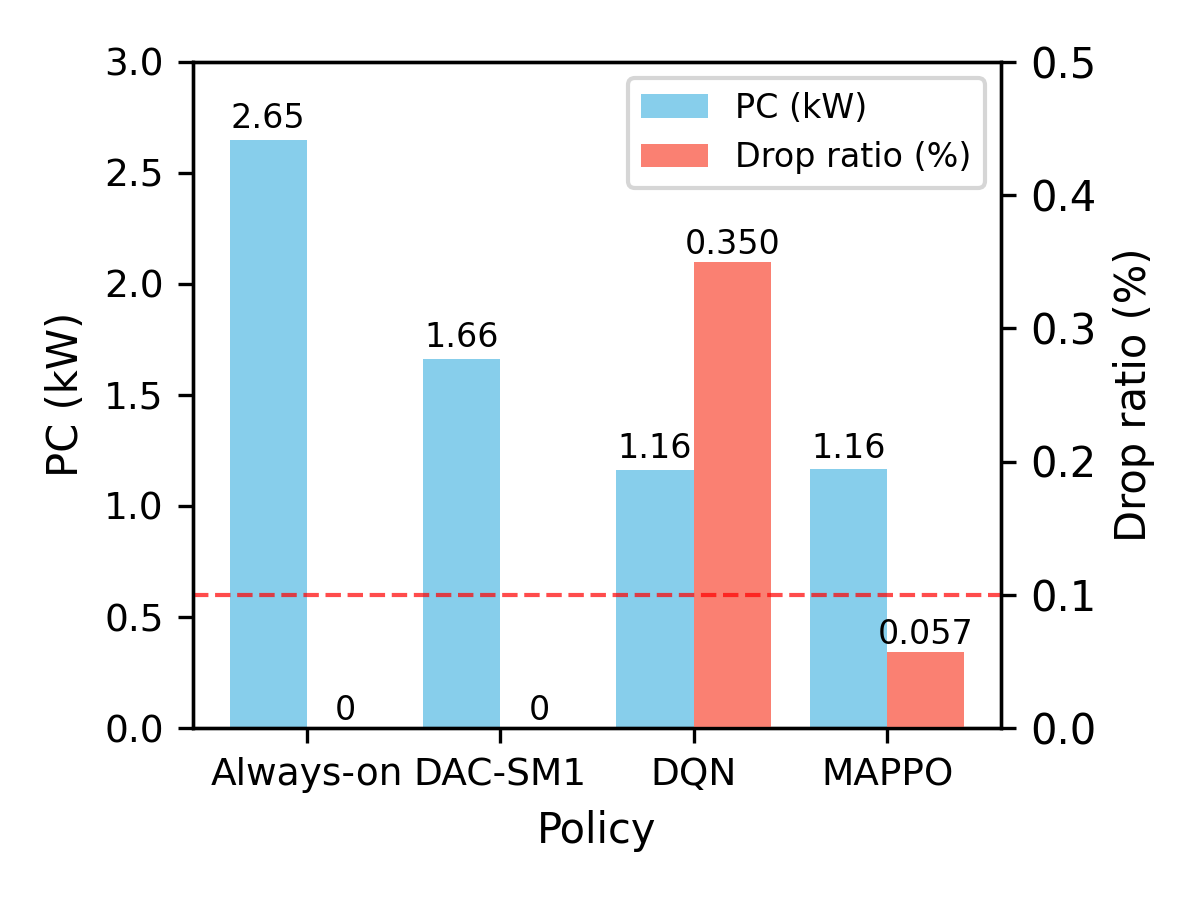}
    \vspace{-1mm}
    \caption{Average PC and drop ratio. \label{fig:kpi}}
\end{minipage}
\end{figure*}

Fig.~\ref{fig:sm} illustrates the number of APs operating in different sleep modes under DQN and MAPPO over a one-day interval. In both cases, the sleep modes are dynamically adjusted according to traffic conditions: during high-traffic daytime hours, a larger number of APs are activated to meet UE demand. However, during low-traffic periods, MAPPO drives more APs into SM3 for energy savings, while keeping more APs active than DQN to maintain service continuity. The daily variation of the total demand rate shown in Fig.~\ref{fig:traffic} reflects the dynamic traffic model described in Section~\ref{sec:traffic}. Fig.~\ref{fig:ant} presents the time-varying average number of active antennas under different policies. MAPPO closely tracks traffic dynamics, with the number of active antennas varying from as few as two to nearly eight, demonstrating a wide adjustment span and strong responsiveness to demand fluctuations. In contrast, both DQN and \emph{DAC-SM1} show limited antenna adjustment ranges. DQN struggles to learn effective control policies under a large action space due to its value-based nature.

The performance of each algorithm is evaluated using network PC and the average drop ratio over all UEs, where the definition of the per-UE drop ratio is given in Section \ref{sec:traffic}. The weekly average performance is presented in Fig.~\ref{fig:kpi}. MAPPO demonstrates superior energy efficiency compared to the non-learning baselines, achieving a 56.23\% reduction in PC relative to the \emph{Always-on} policy and a 30.12\% reduction compared to \emph{DAC-SM1}. These energy savings are obtained with only a negligible increase in the drop ratio, which remains well below the constraint indicated by the red dashed line. In contrast, while DQN achieves a comparable level of PC reduction, it incurs a substantially higher drop ratio, revealing its inability to maintain quality of service under dynamic traffic conditions. As observed from Fig.~\ref{fig:sm} and Fig.~\ref{fig:ant}, although DQN also exhibits some adaptation behavior, this aggregate view does not capture the individual AP dynamics. In practice, the coordination among APs under DQN is less consistent. While DQN can react to traffic variations at a coarse level, it fails to learn stable and cooperative control policies under the dynamic and high-dimensional action space of the multi-agent environment. In contrast, MAPPO allows each AP to make independent decisions while maintaining implicit coordination through the shared critic, resulting in more adaptive and efficient control behavior.
\section{Conclusions}
This paper investigates the challenge of minimizing PC in CF mMIMO networks through the joint optimization of antenna re-configuration and ASMs selection under dynamic traffic conditions. We propose a CTDE MAPPO-based algorithm to learn effective control policies. Simulation results demonstrate that the MAPPO-based approach achieves a 56.23\% reduction in PC compared to the \emph{Always-on} baseline and a 30.12\% reduction relative to \emph{DAC-SM1}, with only a slight increase in drop ratio. Moreover, compared to DQN, MAPPO achieves a significantly lower drop ratio under similar PC levels. These results highlight that the proposed algorithm can capture traffic variations and adjust its actions more effectively, enabling energy-efficient and reliable operation in CF mMIMO networks.

\section*{Acknowledgment}

This work was supported by Swedish Innovation Agency Funded (VINNOVA) through the SweWIN center (2023-00572).

\bibliographystyle{IEEEtran}
\bibliography{references}

\end{document}